\RequirePackage{fix-cm}
\documentclass[twocolumn]{svjour3}          
\smartqed  
\usepackage{graphicx}
\usepackage{epsfig}
\usepackage{amsmath}
\usepackage{mathrsfs}
\usepackage{color}
\usepackage{palatino}
\usepackage{latexsym}
\usepackage{natbib}
\usepackage{lscape}
\usepackage{amssymb}

\journalname{Soft Computing}
\begin{document}

\title{A new flower pollination algorithm for equalization in synchronous DS/CDMA multiuser communication systems
}


\author{Luis M. San-Jos\'e-Revuelta         \and
        Pablo Casaseca-de-la-Higuera 
}


\institute{L.M. San-Jos\'e-Revuelta \at
            ETSI Telecomunicaci\'on, Universidad de Valladolid, 47011 Valladolid, Spain\\
              \email{lsanjose@tel.uva.es}           
           \and
           P. Casaseca-de-la-Higuera \at
           ETSI Telecomunicaci\'on, Universidad de Valladolid, 47011 Valladolid, Spain\\
            \\
            School of Computing, Engineering and Physical Sciences, University of the West of Scotland, Paisley Campus, High Street, Paisley, PA1 2BE, Scotland, UK \\
              \email{jcasasec@tel.uva.es}
           }

\date{This is the final manuscript (post-print) accepted and published in Springer Soft Computing (2020) 24:13069–13083. Digital Object Identifier: 10.1007/s00500-020-04725-x. (c) Springer-Verlag GmbH Germany, part of Springer Nature 2020.}

\maketitle

\begin{abstract}
This work proposes a modified version of an emerging nature-in\-spired technique, named Flower Pollination Algorithm (FPA), for equalizing digital multiuser channels. This equalization involves two different tasks: 1) estimation of the channel impulse response, and 2) estimation of the users' transmitted symbols. The new algorithm is developed and applied in a Direct-Sequence / Code-Divi\-sion Multiple-Access (DS/CDMA) multiuser communications system. Important issues such as robustness, convergence speed and population diversity control  have been in deep investigated. A method based on the entropy of the flowers' fitness is proposed for in-service monitoring and adjusting population diversity. Numerical simulations analyze the performance, showing comparisons with well-known conventional  multiuser detectors such as  Matched Filter (MF), Minimum Mean Square Error Estimator (MMSEE) or several Bayesian schemes, as well as with other nature-inspired strategies. Numerical analysis shows that the proposed algorithm enables transmission at higher symbol rates under stronger fading and interference conditions, constituting an attractive alternative to previous algorithms, both conventional and nature-inspired, whose performance is frequently sensible to near-far effects and  multiple-access interference problems. These results have been validated by running hypothesis tests to confirm statistical significance.

\end{abstract}

\section{Introduction}
\label{sec_intro}

Nowadays communication systems face up the problem of high transmission rates over limited bandwidth channels, with an increasing number of users. In such scenarios, efficient radio channel sharing techniques have deserved special attention in the scientific community during last years. Direct Sequence / Code-Division Multiple-Access (DS/CDMA) is one of the most widely used and studied channel sharing methods. Currently, this technology is implemented in many real wireless communication systems, such as LTE, UMTS, CDMA2000, IS-95 and even in global positioning applications (US GPS, European Galileo and Russian GLONASS). Besides, CDMA is pre\-sent in 5G systems in combination with other technologies as well as in cordless phones operating in the 900 MHz, 2.4 GHz and 5.8 GHz bands, in IEEE 802.11b 2.4 GHz wifi, and in radio controlled model automotive vehicles.  In modern systems where security is a key design aspect, DS/CDMA has been proposed combined with both  chaotic spreading sequences \citep{Rahnama13} and with MIMO techniques \citep{Mehrizi17} to decrease the level of interferences, improving, this way, other sharing schemes.

Due to its extensive application and use, DS/CDMA systems constitute a well-known and widely studied problem, thus allowing direct comparison to previously developed multiuser detectors.

Since current CDMA systems allow transmission at very high rates, intersymbol and multiaccess interferences (ISI and MAI) must be 
 controlled so that performance does not degrade \citep{Proakis98}. These interferences, if not limited, will surely drop system's performance. To address this problem, conventional multiuser detectors (MUDs) make use of filters matched to the codeword of the user of interest. However, this scheme is only optimum when all received codewords are independent, which rarely occurs in real applications and performance notably decreases, especially if near-far effects are also present. In 1998, S. Verd\'u  found that the joint extraction of the users transmitted sequences could mitigate this problem \citep{Verdu98}. However, the optimum MUD relying on the maximum likelihood (ML) criterion has a complexity that is exponentially proportional to the number of system users, making it unfeasible in real scenarios. Therefore, many authors have paid attention to the development of suboptimal schemes that can be implemented on real systems.

An important branch of methods are those based on meta-heuristics, specially, nature-inspired approaches. These have been successfully tested for up to 2,000 dimensions for large scale optimization \citep{Agarwal14}. Among these nature-inspired approaches the following can be highlighted: Genetic Algorithms (GAs) were the first and simplest methods. \cite{Juntti97} proposed a synchronous DS/CDMA MUD based on GAs. Its main drawback was the requirement of good estimates of the first transmitted symbols.  A variant of the GA was proposed by \cite{Yen00}, whose performance is close to the optimum thanks to a local search algorithm introduced before the GA. This idea of adding modules to the standard GA was also used by \cite{Ergun00}. In this case a multistage detector is integrated into the GA in order to speed up convergence. Later, \cite{Yen04} studied the  asynchronous case taking into account  the effect of the surrounding symbols of the other system users.

Other 
 works include \citep{Maradia09}, which focuses on selective Rayleigh fading channels estimated using a GA-based detector, and the approach by \cite{Dong04}, which studies both bit-error-rate (BER)  and near-far effect performances. Some other worth-mentioning recent approaches using GAs are \citep{Yen01,Shayesteh03,Tan10,Tan10b,Nooka13,Khan15,Huang17}.

Apart from GAs, several other nature-inspired methods have also been applied to multiuser detection. For instance:  tabu search (TS) has been proposed in \citep{Tan04,Driouch10,Datta10,Srinidhi11}. A relevant simulated annealing (SA)-based strategy was proposed by  \cite{Tan04}. Hybrid methods combining SA with particle swarm optimization, genetic algorithms and a Hopfield neural network have been proposed in \citep{Gao09,Yao11,Tan13}, respectively. In methods such as TS or SA, a single solution is obtained throughout search, therefore the quality of the solution highly depends on initial guesses.

Other nature-inspired methods are the more advanced particle swarm optimization (PSO) \citep{Oliveira06,Soo07,Wang14,Arani13,Kaur16}, cat swarm optimization (CSO) \citep{Pradhan12,Sohail17,Balamurugan18}, the cuckoo search algorithm (CSA) \citep{Agarwal14,Yang14}, ant-colony algorithms (ACO)  \citep{Hijazi04,Lain07,Xu07,Marinello12} or the Flower Pollination Algorithm (FPA) \citep{Yang12,Balasubramani14,Pant17,Abdel17b,Abdel19}, among others. See \citep{Larbi14} for an interesting list of references using heuristic methods in CDMA communication problems.

The existing literature reveals that the main drawbacks in solving real-world problems found in these metaheuristics are: computational complexity, slow convergence speed and suboptimal convergence to local minima, mainly due to the difficulty to maintain population diversity while iterations run, and the impossibility to detect and escape from these suboptimal solutions \citep{Tan04,Agarwal14,Larbi14,Arora17,Sohail17}.

In this paper we focus on the development of a new  nature-inspired algorithm, based on the standard Flower Pollination Algorithm (FPA) --initially proposed by \cite{Yang12}-- and its application to channel estimation and symbol detection in DS/CDMA environments. Among the main advantages of FPA we find: easy implementation, shorter computation time, broad flexibility, robust performance, intuitive structures and guarantee of fast conver\-gence \citep{Yang12,Abdel14,Dubey15,Nigdeli16,He17,Abdel19}. In fact, FPA has already been successfully used in complex applications such as: constrained global optimization \citep{Abdel14}, economic dispatch \citep{Prathiba14,Dubey15}, assembly sequence optimization \citep{Mishra19}, analysis of micro-CT scans \citep{Kowalski19}, simulation of photovoltaic systems \citep{Alam15}, EEG-based person identification \citep{Rodrigues16}, knapsack problems \citep{Abdel18,Abdel18c}, the spherical traveling salesman problem \citep{Zhou19}, ill-conditioned sets of equations \citep{Abdel16},  quadratic assignment \citep{Abdel17}, antenna positioning \citep{Dahi16}, wireless sensor optimization \citep{Sharawi14}, or sizing optimization of truss structures \citep{Bekdas15}, among others. Recently, FPA has also been applied in hybrid proposals with other nature-inspired algorithms, such as tabu search, simulated annealing or particle swarm optimization  \citep{Lenin14,Abdel15,Hezam16,Abdel16}. A detailed review of FPA applications can be found in \citep{Balasubramani14,Chiroma15,Pant17}.

Our proposal introduces some modifications to the standard FPA in order to: i) improve search convergence by monitoring and controlling population diversity using the entropy of the population fitness, and 2) reduce its complexity by carefully adjusting the probability of each pollination type (local/ global) and considering different applications of these pollinations to certain parts of the potential solutions --see section \ref{secc_fpa}.

To the best of our knowledge, no previous attempts to make use of FPA for joint fading estimation and symbol detection in communications problems have been published.

The rest of the paper is organized as follows: section 2 explains the joint channel estimation and symbol detection problem in a DS/CDMA communications system. Basic concepts and notation are there presented, along with the proposed fitness function to be used in the FPA. Section 3 develops the proposed FPA and shows how population diversity is controlled using the population fitness entropy. Next, section 4 shows the numerical results with emphasis to comparison with both conventional and nature-inspired multiuser detectors.  Finally, conclusions and future research are outlined in section 5.

\section{DS/CDMA multiuser communication channel model}
\label{dscdma}

This section describes the channel model used, where $U$ binary symbol sequences are simultaneously transmitted over a shared radio channel. Each user makes use of a private normalized modulation signature from the set $\{s_i(t)\}_{i=1}^U$. Channel response is considered to be completely characterized by a set of flat-fading coefficients, and an additive white zero-mean Gaussian noise (AWGN). A perfect synchronization of all signals is assumed.

User $i$ transmits an $F$-length sequence $d_i(n)$ of sta\-tis\-ti\-cal\-ly-independent symbols that modulate the codeword $s_i(t)$. This way, the $i$th user transmits the following signal
\begin{eqnarray}
    x_i(t) = \sum_{n=0}^{F-1} d_i(n) s_i(t-nT)
\label{xit}
\end{eqnarray}
where $d_i(n)$ are the users' transmitted symbols and $T$ is the symbol period. User signatures (or codewords) are obtained as
\begin{eqnarray}
    s_i(t)=\sum_{\ell=0}^{N-1} s_{i,\ell} \gamma (t-\ell T_c)
\label{sit}
\end{eqnarray}
where ${\bf s}_i=(s_{i,0},...,s_{i,N-1})^T$ stands for the $i-th$'s user signature, $T_c=T/N$ is the chip period ($N$ is known as {\it processing gain})  and $\gamma(t)$ denotes a chip waveform with normalized energy. As a consequence, the spectrum width is spread by a factor $N$, de-sensitizing, this way, the original narrowband signal to some potential channel degradation and interference \citep{Proakis98}. As we are considering a  synchronous system, the signal at the receiver input is
\begin{eqnarray}
    r(t) &= \sum_{i=1}^U r_i(t) + g(t)     \qquad    0 \leq t \leq T_F
\label{rt}
\end{eqnarray}
where $g(t)$ denotes an AWGN, which is not correlated with symbols $d_i(n)$, $T_F$ represents the frame duration and $r_i(t)$ is
\begin{eqnarray}
    r_i(t) = \sqrt{E_i} \sum_{n=0}^{F-1}  a_i(n) d_i(n) s_i(t-nT)
\label{rt2}
\end{eqnarray}
$E_i$ denotes the bit energy of the $i$th user, and $a_i(n)$ is the flat-fading coefficient of the $i$th user. In this work, a non-stationary channel is considered, with a time variation  model defined in \citep{Yen01}, where fading coefficients $a_i(n)$ change with time as a function of a Doppler frequency, $f_d$, as
\begin{eqnarray}
    a_i(n+1) = \alpha \cdot a_i(n) + \nu
\end{eqnarray}
with $\alpha = e^{-2\pi f_d T}$ and $\nu$ representing an AWGN component.

The main aim of our algorithm --the joint channel estimation and symbol detection task-- can be seen in Eq. (\ref{rt2}), where both the users' data symbols as well as the flat-fading  coefficients must be estimated.

\begin{figure*}[htb]  
\centerline{\psfig{file=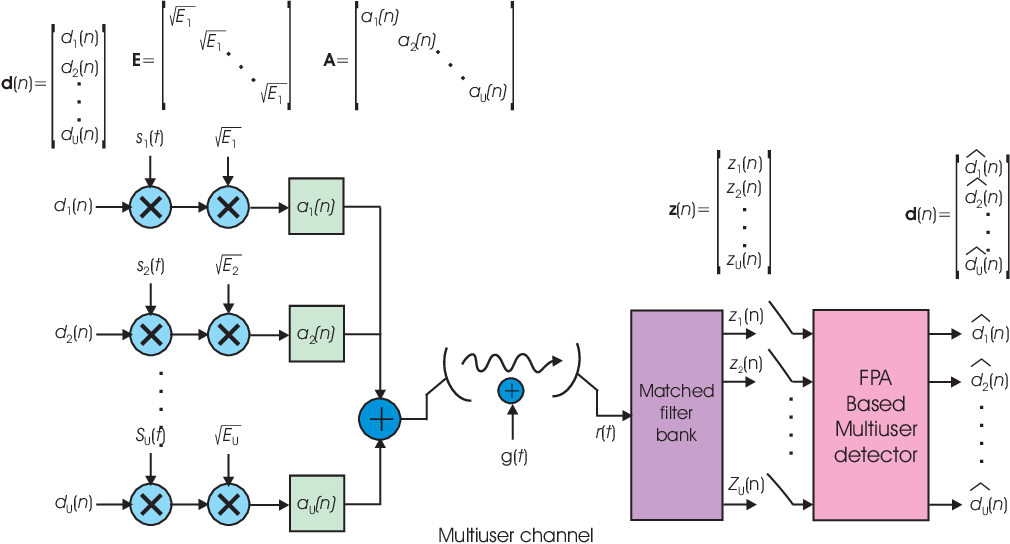,width=15cm}}
\caption{Multiuser DS/CDMA system model with $U$ active users. $E_i$: bit energy of user $i$, $d_i(n)$: symbol sequence transmitted by user $i$, $s_i(t)$: codeword of user $i$, $\hat{d}_i(n)$: estimate of $n$th symbol transmitted by user $i$. $a_i$: channel flat-fading coefficient of user $i$.} 
\label{fig_channel}
\end{figure*}

The first step in the receiver consists of a bank of filters  matched to the users' codewords, just after sam\-pling the received signal at $1/T$ rate --see Fig. \ref{fig_channel}.

In order to estimate the vector of transmitted symbols ${\bf d}(n)=[d_1(n),\dots,d_U(n)]^T$, we have followed ideas in \citep{Yen01}, where estimation is presented as a maximization problem, and the output of the matched filters' bank, ${\bf z}(n)$, is obtained as
\begin{eqnarray}
    {\bf z}(n)=[z_1(n),\dots,z_U(n)]^T = {\bf R} {\bf A}(n) {\bf E} {\bf d}(n) + {\bf g}(n)
\end{eqnarray}
with ${\bf R}$ being the $U \times U$ cross-correlation matrix of users' codewords, ${\bf A}(n)=\mbox{diag}(a_1(n),\dots,a_U(n))$, ${\bf E}=\mbox{diag}(\sqrt{E_1},\dots,\sqrt{E_U})$, ${\bf d}(n)=[d_1(n),\dots,d_U(n)]^T$ and ${\bf g}=[g_1(n),\dots,g_U(n)]^T$.

Considering vector ${\bf z}$, it can be demonstrated  that the log-likelihood conditional pdf given both the fading coefficients' matrix ${\bf A}(n)$ and the users' data vector ${\bf d}(n)$, is \citep{Fawer95}
\begin{multline}
    {\mathcal L}\left( {\bf A}(n), {\bf d}(n)  \right)  =  2 \Re \left\{ {\bf d}(n)^T {\bf E} [{\bf A}(n)]^* {\bf z}(n)  \right\}      \\
        -  {\bf d}(n)^T {\bf E} {\bf A}(n) {\bf R} [{\bf A}(n)]^* {\bf E} {\bf d}(n)
\label{metrica}
\end{multline}
with ``$\Re$'' and ``*'' representing the real part of a complex magnitude, and the complex conjugate operator, respectively. Hence, estimates of the matrix with the fading coefficients and the vector of data symbols are obtained as
\begin{eqnarray}
    (  \widehat{{\bf A}(n)},  \widehat{{\bf d}(n)}   )  =    \arg \max_{{\bf A}(n),{\bf d}(n)}  \left\{ {\mathcal L} \left( {\bf A}(n),{\bf d}(n) \right)  \right\}
\label{maxprob}
\end{eqnarray}

On the other hand, time variation of fading coefficients is supposed to be slow enough so as to consider fading response to be constant within each symbol period. Besides, fadings from different users are considered independent.

Notice that this receiver works at symbol rate $1/T$. However, it is also possible to develop an algorithm operating at chip rate, $1/T_c = N/T$. A brief introduction to this approach is developed in Appendix \ref{Apendice_A}.

\section{Flower pollination algorithm (FPA) description}
\label{secc_fpa}

\subsection{Basic concepts and standard FPA}

The standard FPA \citep{Yang12} is based on modelling the natural pollination process in nature. This process is necessary for reproduction in those plants with flowers (angiosperms), which are about 80\% of all plant species. The pollination process transfers pollen from some flowers to others, with the aim of reproduction. Two types of pollination can be considered: biotic (pollen is transferred by insects or animals) and abiotic (the wind or the rain carry out the process of pollination).

Two different types of pollination can be considered: local  (or {\em auto-pollination}) and global (or {\em cross pollination}). In the later, pollen is transferred from the flower of another plant; while, in local pollination, pollen comes from the same flower, or flowers, belonging to the same plant, usually without the action of a pollinator.

Global pollination is normally biotic and pollen travels long distances due to the action of pollinators such as birds, insects, bees or bats, whose travels can be modeled using a L\'evy distribution \citep{Pavlyukevich07}.

According to \citep{Yang12}, the basic process of pollination  can be idealized considering that: (i)  biotic cross pollination is {\em global} pollination, (ii) pollinators use L\'evy flights to carry pollen, (iii) auto-pollination (abiotic) is a {\em local} pollination, and (iv) each type of pollination is selected using a variable $P_c \in [0,1]$, known as {\em probability of change}. For simplicity, we will assume that each plant has just one flower and each flower has only one gamete of pollen. This way, {\em pollen},  {\em flower} and {\em  plant} are equivalent concepts. In our channel estimation and symbol detection problem, each flower or plant is represented by a solution vector ${\bf x}_i[t]$:
\begin{eqnarray}
    {\bf x}_i[t] &=& [{\bf d}(t),{\bf a}(t)]   \nonumber \\
                 &=& \left[d_{i,t}(1),d_{i,t}(2),\dots \right. \nonumber \\
  & & \left.  \dots , d_{i,t}(U),a_{i,t}(1),a_{i,t}(2),\dots , a_{i,t}(U)  \right]
\label{ec_49}
\end{eqnarray}
where index $i$ indicates the $i$th flower from a population of $NumFl$ $(1 \leq i \leq NumFl)$ flowers, and $t$ stands for current generation. In Eq. (\ref{ec_49}) we se that each solution vector  consists of two parts: first, an estimation of the $U$ symbols transmitted by users  --left part of ${\bf x}_i[t]$--, and, secondly, the estimation of the $U$ flat-fading channel response coefficients --on the right. Population will consist of a total of $NumFl$ possible solutions, i.e.
\begin{multline}
\mbox{Population}[t] =   \\
\left\{
\begin{array}{l}
{\bf x}_1[t] = \left[ d_{1,t}(1),d_{1,t}(2),\dots , d_{1,t}(U), \dots   \right. \\
     \hspace{3cm} \left.   a_{1,t}(1),a_{1,t}(2),\dots , a_{1,t}(U)  \right]  \\
   {\bf x}_2[t] = \left[ d_{2,t}(1),d_{2,t}(2),\dots , d_{2,t}(U),\dots   \right.  \\
    \hspace{3cm} \left. a_{2,t}(1),b_{a,t}(2),\dots , a_{2,t}(U)  \right] \\
\vdots  \\
    {\bf x}_{NumFl}[t] = \left[ d_{NumFl,t}(1),d_{NumFl,t}(2),\dots  \right. \\
   \hspace{1.4cm}  \dots d_{NumFl,t}(U), a_{NumFl,t}(1),\dots \\
   \hspace{1.4cm}   \left.  \dots  a_{NumFl,t}(2),\dots , a_{NumFl,t}(U)  \right]
\end{array}
\right.
\label{ec_50}
\end{multline}
%

The standard FPA has two main steps: global pollination and local pollination. The first one aims at modelling the long travels of insects with the aim of guaranteeing optimal reproduction; this is formulated as
 \begin{eqnarray}
    {\bf x}_i[t+1] =  {\bf x}_i[t]  +  \gamma {\bf L} ( {\bf b}^* - {\bf x}_i[t])
\label{ec_51}
\end{eqnarray}
Where ${\bf x}_i[t]$ represents the $i$th pollen at iteration $t$, ${\bf b}^*$ is the fittest found solution until iteration $t$, $\gamma$ represents the scale factor (or {\em jump}), and ${\bf L}$ is the {\em pollination strength} or {\em force}, which is basically a $2U$-dimensional step-size. A L\'evy flight \citep{Pavlyukevich07}, ${\bf L}$, is used for modeling long travels of pollinators. A Levy flight is a random movement process similar to a random walk, but with the particularity that moves have a random length generated from a heavy-tailed probability distribution \citep{Dahi16}. This parameter is computed as
\begin{eqnarray}
    {\bf L} =      \frac{\beta \Gamma (\beta) \sin(\pi \beta /2)  }{ \pi } \frac{1}{s^{1+\beta}}, \qquad s \gg s_0>0
\label{ec_52}
\end{eqnarray}
with  $\Gamma$ representing the standard gamma function  (defined as $\Gamma(x) = \int_0^\infty u^{x-1} e^{-u} du$), and  $\beta$ a tuning parameter set to 1.5 as suggested in \citep{Yang12}. Notice that Eq. (\ref{ec_52}) can only be applied for large steps $s \gg 0$. 
Step $s$ is obtained using two random gaussian numbers, $U\sim N(0,\sigma^2)$ and $V \sim N(0,1)$, as
\begin{eqnarray}
    s =      \frac{U}{|V|^{1/\lambda}}
\label{ec_s}
\end{eqnarray}
and standard deviation $\sigma$ in $U$ is calculated using
\begin{eqnarray}
    \sigma = \left\{     \frac{\Gamma(1+\lambda)}{\lambda \Gamma [ (1+\lambda)/2  ]}    \frac{\sin(\pi \lambda /2)}{2^{(\lambda -1)/2}}         \right\}^{1/\lambda}
\label{ec_sigma}
\end{eqnarray}

Finally, parameter $\gamma$ is set to 0.1 as suggested and used in \citep{Yang12,Abdel18}, and $\lambda$ is set to 1.0   after several heuristic trials in the range $[0.7 - 2]$, taking into account that \cite{Rycroft05} proved that the best value of $\lambda$ is between 0.75 and 1.95.

On the other hand, local pollination is formulated as:
\begin{eqnarray}
    {\bf x}_i[t+1] =  {\bf x}_i[t]  +  \epsilon ( {\bf x}_j[t] - {\bf x}_k[t])
\label{ec_53}
\end{eqnarray}
Where ${\bf x}_j[t]$ and ${\bf x}_k[t]$ represent flowers $j$ and $k$ of generation $t$; i.e., two solution vectors randomly selected from the population. These flowers belong to the same plant, simulating local pollination. Random variable $\epsilon$ is obtained using a uniform distribution in $[0,1]$.

As mentioned before, in nature, pollination occurs on both  local and global scale. In practice, it is more likely that flowers next to each other will be pollinated by pollen from nearby flowers, and not by those that are further away. In the FPA, probability of change ($P_c$) is used to switch between global and local pollination. In our implementation $P_c$ denotes the probability of global pollination, with global pollinations defined as complementary events. Thus, probability of local pollination is $(1-P_c)$. Following ideas in \citep{Draa15} we started with $P_c = 0.2$, and, after several simulations, we decided to use $P_c = 0.35$, a value that has proved to achieve good results in most cases.

\subsection{Diversity monitoring and control}
\label{sec_divmon}

An important novelty of our FPA is the introduction of an scheme for fine-tuning the probability of change ($P_c$) in order to achieve and maintain diversity within the population of flowers. A well-known drawback of evolutionary  algorithms is their tendency to converge to suboptimal solutions \citep{Agarwal14,Larbi14}. This is in part due to the use of selection strategies relying directly on fitness values. As a consequence, a reduced set of repeated  best individuals will compose the population, resulting in a very low diversity.
In this paper, we propose to use a probability of change that depends on the Shannon entropy of the population fitness, which can be computed as
\begin{eqnarray}
    {\mathcal H}({\mathcal P}[k]) = - \sum_{i=1}^{NumFl} {\mathcal L}_i^*[k] \log {\mathcal L}_i^*[k]
\label{entropia}
\end{eqnarray}
with  ${\mathcal L}_i^*[k]$ being the normalized fitness of flower  ${\bf x}_i$, i.e. ${\mathcal L}_i^*[k] = {\mathcal L}_i[k] / (\sum_{j=1}^{NumFl} {\mathcal L}_j[k]  )$. Fitness ${\mathcal L}_i[k]$ is calculated before normalization using Eq. (\ref{metrica}). This way, when fitness values are all of them very similar, with small dispersion,  ${\mathcal H}({\mathcal P}[k])$ becomes high, and $P_c$ is increased in order to boost global pollination through Eq. (\ref{ec_51}). Thus, diversification is performed. On the other hand, a small value of ${\mathcal H}({\mathcal P}[t])$ reveals a high diversity within population, and $P_c$ is decreased so as to promote local pollination through Eq. (\ref{ec_53}).

At this point, it is important to notice that entropy will be high both during first iterations (fitness values are low since they are initial estimates randomly generated) and last iterations (if convergence has been achieved most of the flowers will encode similar solution estimates representing good estimates of the solution with a high associated fitness value). This situation is easy to manage since we know if the algorithm is in the first or last iterations.

Another particularity is that pollination operators given by Eqs. (\ref{ec_51}) and (\ref{ec_53}) are now applied individually to each part of flower ${\bf x}_i[t]$ --see Eq. (\ref{ec_49}). Global pollination will affect the right-most part of ${\bf x}_i[t]$ (regarding flat fading coefficients) with higher probability during the first iterations. After an initial period of symbols, this part will be affected almost only by local pollination. On the other hand, the left side of ${\bf x}_i[t]$ (encoding users' transmitted data) will be affected following the standard FPA principles. The reason for this is that channel coefficients vary slowly with time and can be considered constant along several symbol periods. Once they are estimated, only a fine-tuning of them is necessary. Transmitted symbols are considered independent between different symbol periods, so FPA must be initialized with randomly generated symbol estimates, for every transmitted symbol.

In order to perform this adjustment, flower coding is divided into two zones of different pollination. The part of the fading coefficients ({\bf a}(n)) does not need to be re-initialized when the algorithm is run in each symbol period. On the contrary, the part encoding transmitted symbols (${\bf d}(n)$) is randomly re-initialized at the beginning of each symbol period --see structure of ${\bf x}_i[t]$ in Eq. (\ref{ec_49})--.

According to this, and taking into account that ${\bf x}_i[t] = [ {\bf d}(n) | {\bf a}(n) ]$, local pollination is performed with a lower $\gamma_a$ for the flat-fading coefficients part, and with $\gamma_b$ ($\gamma_b > \gamma_a$) for the ${\bf d}(n)$ part of ${\bf x}_i[t]$. Specifically, $\gamma_b(0)=0.1$ and remains constant with time, while $\gamma_a(0)=0.05$ and decreases $10\%$ every $MaxIter /4$.

The proposed FPA can be summarized using the following steps:

\begin{enumerate}
\item The iteration counter is initialized to 0,  and an initial population of $NumFl$ flowers ({\bf x}) is established with randomly generated values. The best solution ${\bf b}^*$ in this initial population is found (the one with the lowest fitness value). Probability of change $P_c(0)$ is set to 0.35.

 \item A random value $r \in [0,1]$ is calculated. If $r < P_c$ then go to step 3 (global pollination), otherwise, go to step 4 (local pollination).

\item  Jump vector {\bf L} is estimated using the L\'evy distribution given by Eq. (\ref{ec_52}). Then, global pollination is applied according to:
 \begin{eqnarray}
    {\bf d}_i[t+1] =  {\bf d}_i[t]  +  \gamma_d {\bf L} ( {\bf d}^* - {\bf d}_i[t])  \label{ec_51a}  \\
    {\bf a}_i[t+1] =  {\bf a}_i[t]  +  \gamma_a {\bf L} ( {\bf a}^* - {\bf a}_i[t])  \label{ec_51b}
\end{eqnarray}
    with $\gamma_d > \gamma_a$, and ${\bf d}^*$ and ${\bf a}^*$ representing the corresponding part in ${\bf b}^*$ (current best solution), i.e. ${\bf b}^* = [ {\bf d}^* | {\bf a}^* ]$.

\item $\epsilon$ is drawn from a uniform distribution in the interval [0,1]. Flowers $j$ and $k$ are randomly selected from the current population ($1 \leq j,k \leq NumFl$, $j \neq k$). Local pollination is applied using Eq. (\ref{ec_53}).

\item The fitness value for each flower is calculated using Eq. (\ref{metrica}). If the new fitness value for each flower is better than the previous one, then flower {\bf x} is updated. If not, the solution from previous iteration is kept.

\item Entropy of population fitness is evaluated using Eq. (\ref{entropia}) and $P_c$ is accordingly modified:

\begin{itemize}
    \item ${\mathcal H}({\mathcal P}[t]) \uparrow \quad \Rightarrow$ low dispersion $\Rightarrow \quad \uparrow P_c$ $\Rightarrow$ $\uparrow$ exploration.

    \item ${\mathcal H}({\mathcal P}[t]) \downarrow \quad \Rightarrow$ large dispersion $\Rightarrow \quad \downarrow P_c$ $\Rightarrow$ $\uparrow$ exploitation.
\end{itemize}
        Modifications of $P_c$ are smaller as iteration $t$ increases and are only applied every $MaxIter/10$ iterations.

 \item Once every flower has been analyzed and pollination has been applied, the gamete with the best fitness is selected and assigned to ${\bf b}^*=[ {\bf d}^* |  {\bf a}^*  ]$.

\item The algorithm finishes if the preset maximum number of iterations ($MaxIter$) is reached. If not, go to step 2.

\end{enumerate}

\begin{figure}[!htb]  
\centerline{\psfig{file=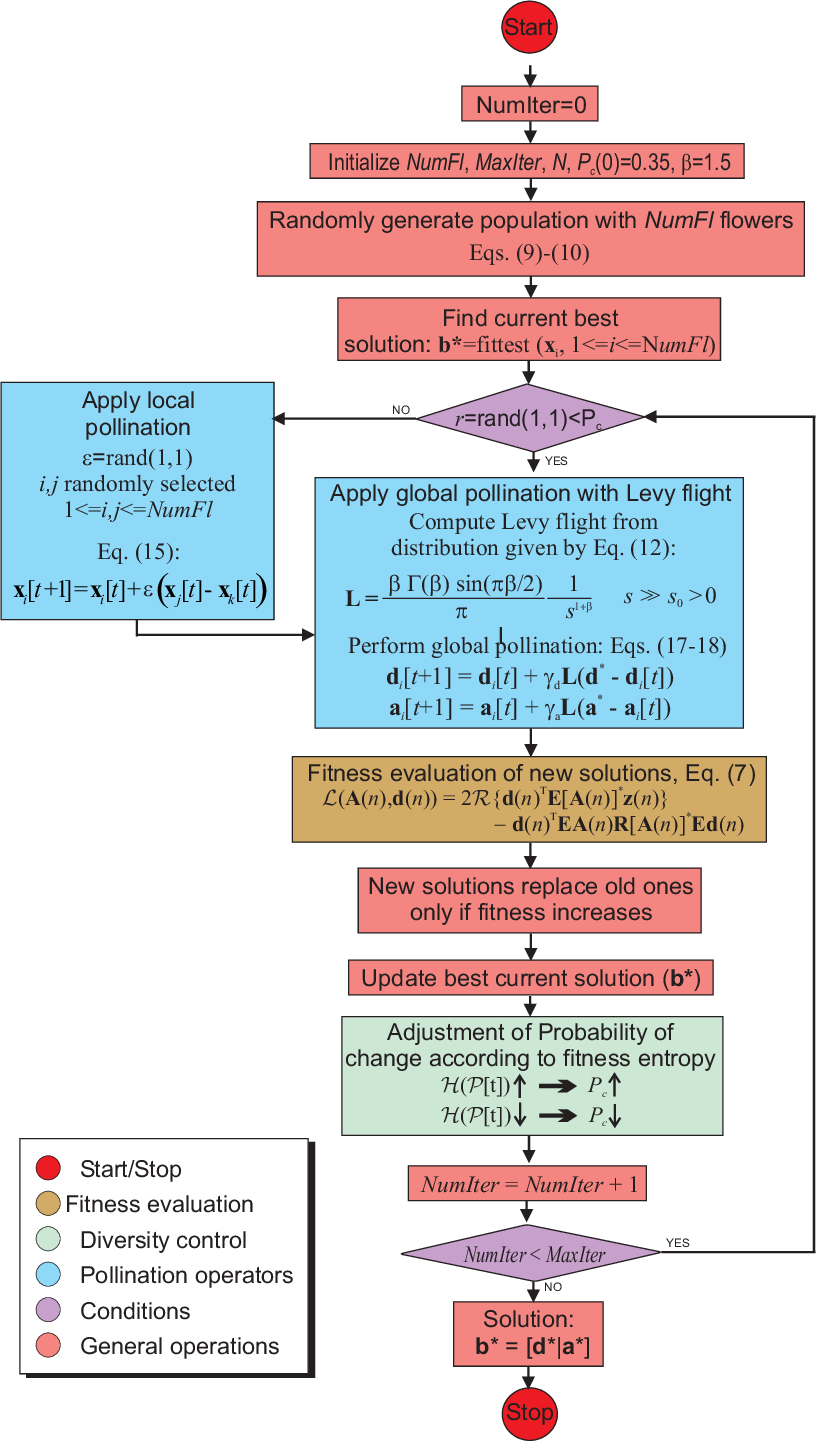,width=9cm}}
    \caption{Flowchart of the proposed FPA (flower pollination algorithm).}  
\label{fig_flowchart}
\end{figure}

Fig. \ref{fig_flowchart} shows a flowchart corresponding to the proposed FPA.

When the algorithm finishes, the flower  with the highest fitness will represent the problem solution, i.e. the estimate of both the symbols' vector {\bf d} and the matrix with fading coefficients {\bf A}, whose diagonal elements are represented in vector ${\bf a}$.

\section{Experiments}
\label{sec_nr}

This section shows the results of the Monte Carlo simulations carried out to evaluate  performance of the proposed FPA-based multiuser detector. Unless otherwise specified, a synchronous scenario with additive white Gaussian noise will be considered. In addition, Binary Phase Shift Keying (BPSK) modulation,  rectangular chip pulse  $\gamma(t)$ -see Eq (2)-, and Gold sequences (with $N=31$ chips) as signatures $s_i(t)$, are used. Besides, time variation is defined by a Doppler frequency $f_d = 1000$ rad/sec. By default, $U$=10 simultaneously transmitting users are considered. The proposed FPA uses {\em MaxIter}$=2,000$ iterations, {\em NumFl}$=25$ flowers, and is implemented with a probability of change that depends on the Shannon entropy of the fitness values  of the population as explained in step 6 of section \ref{sec_divmon}.

\subsection{Bit Error Rate (BER) vs Signal-to-Noise Ratio (SNR)}

\begin{figure}[htb] 
\centerline{\psfig{file=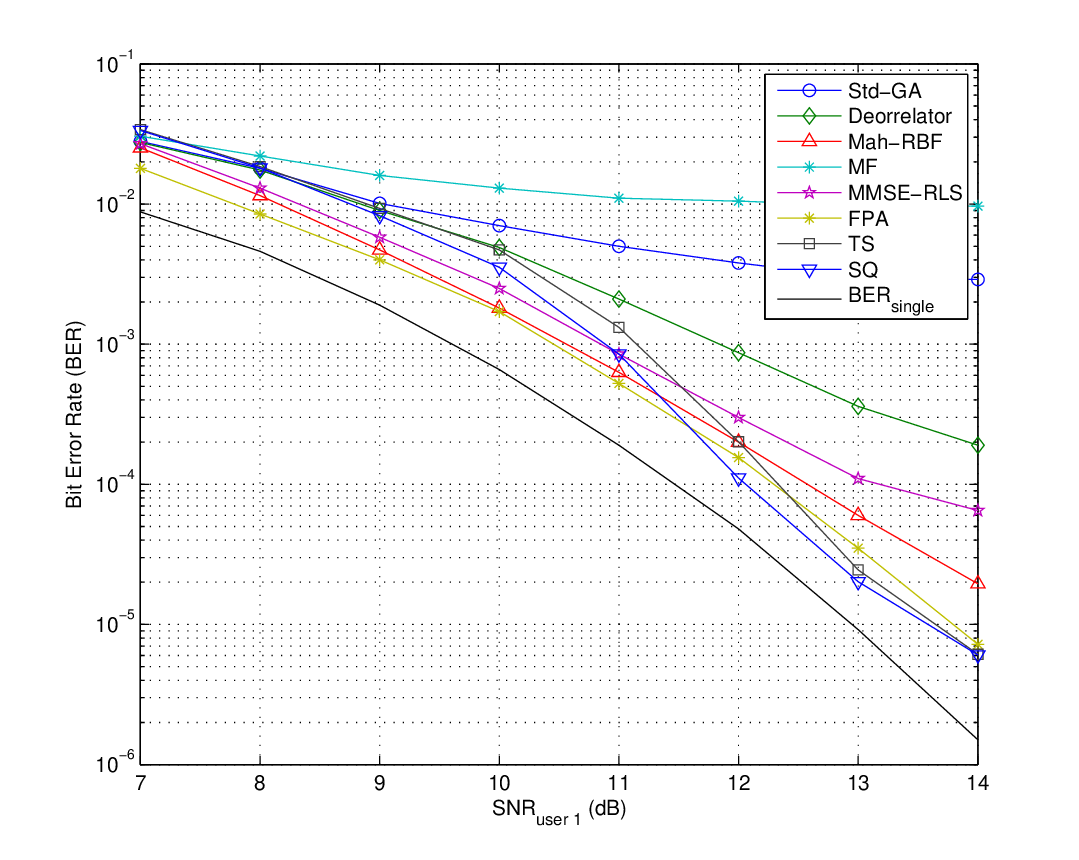,width=9cm}}
\caption{Performance in terms of the probability of bit error (BER) depending on the SNR (dB) for several conventional and nature-inspired multiuser detectors. Near-far distortion: $E_j/E_1$=4 dB for $j \geq 1$. $n_p$: population size of the standard GA.}  
\label{fig_ber_snr1}
\end{figure}

Considering user No. 1 as the user of interest (UOI), Fig. \ref{fig_ber_snr1} shows its bit error probability as a function of the signal quality, which is evaluated in terms of $SNR_{user\_1}$ $=E_1/N_0 \in \{7,\dots,14 \}$ dB, for different conventional (decorrelator RLS \citep{Lim98}, matched filter \citep{Lupas89}, MMSE RLS) and nature-inspired (Standard GA, Mahalanobis RBF  \citep{SanJoseNNSP03}, Tabu Search, Simulated Quenching and proposed FPA) algorithms. The RBF detector is implemented with the Mahalanobis distance as kernel function. Powers of the remaining users are set with a near-far effect of 4 dB (i.e. $E_j/E_1$ = 4 dB, $2 \leq  j \leq U$).

\begin{table*}[htb]  
\centering
\resizebox{18cm}{!}{
\begin{tabular}{|c|c|c|c|c|c|c|c|c|}
\hline
      &    \multicolumn{8}{c|}{ $SNR_{user \hspace{0.1cm} 1}$ (dB)}   \\
\hline
      &  7    & 8  &  9 &  10 & 11  & 12  &  13  & 14  \\
\hline
 Standard GA       &  $0.028 \pm 0.012$   & $0.018 \pm 0.004$  &  $0.0101 \pm 0.002$  &  $0.007 \pm 0.019$  &  $0.005 \pm 0.001$  &  $0.0038 \pm 0.002$  &  $0.003 \pm 0.0014$  &  $0.0029 \pm 0.0009$ \\
\hline
Matched Filter     &  $0.031 \pm 0.009$   & $0.022 \pm 0.009$  &  $0.016 \pm 0.009$  &  $0.013 \pm 0.004$  &  $0.011 \pm 0.002$  &  $0.011 \pm 0.001$  &  $0.009 \pm 0.0012$  &  $0.009 \pm 1.1 \times 10^{-3}$ \\
\hline
Decorrelator RLS   &  $0.027 \pm 0.005$   & $0.017 \pm 0.002$  &  $0.009 \pm 0.001$  &  $0.005 \pm 0.001$  &  $0.002 \pm 0.0010$  &  $0.0009 \pm 1.4 \times 10^{-4}$  &  $3.6 \times 10^{-4} \pm 0.9 \times 10^{-4}$  &  $0.0002 \pm 7 \times 10^{-5}$ \\
 \hline
 Mahalanobis RBF   &  $0.025 \pm 0.005$   & $0.011 \pm 0.0019$  &  $0.0047 \pm 0.0020$  &  $0.0018 \pm 0.0011$  &  $0.0006 \pm 0.0002$  &  $2.0 \times 10^{-4} \pm 1.1 \times 10^{-5}$  &  $6.0 \times 10^{-5} \pm 1.2 \times 10^{-5}$  &  $1.9 \times 10^{-5} \pm 0.18 \times 10^{-5}$ \\
\hline
MMSE-RLS          &  $0.027 \pm 0.010$   & $0.013 \pm 0.005$  &  $0.006 \pm 0.001$  &  $0.0025 \pm 0.0011$  &  $0.0008 \pm 0.0001$  &  $3.0 \times 10^{-4} \pm 7.9 \times 10^{-5}$  &  $1.1 \times 10^{-4} \pm 6.9 \times 10^{-5}$  &  $6.5 \times 10^{-5} \pm 2 \times 10^{-5}$ \\
\hline 
Proposed FPA    &  $0.018 \pm 0.012$   & $0.0085 \pm 0.0011$  &  $0.004 \pm 0.0007$  &  $0.0017 \pm 0.0006$  &  $5.2 \times 10^{-4} \pm 1.3 \times 10^{-4}$  &  $1.5 \times 10^{-4} \pm 1.6 \times 10^{-5}$  &  $3.5 \times 10^{-5} \pm 0.8 \times 10^{-6}$  &  $7.2 \times 10^{-6} \pm 0.9 \times 10^{-6}$ \\
\hline 
Tabu Search       &  $0.034 \pm 0.005$   & $0.018 \pm 0.003$  &  $0.0093 \pm 0.002$  &  $0.005 \pm 0.0008$  &  $0.0013 \pm 0.0006$  &  $0.0002 \pm 2 \times 10^{-5}$  &  $4.4 \times 10^{-5} \pm 2.2 \times 10^{-6}$  &  $6.1 \times 10^{-6} \pm 1.4 \times 10^{-6}$ \\
\hline
Simulated Quenching & $0.033 \pm 0.015$  & $0.018 \pm 0.011$  &  $0.008 \pm 0.0015$  &  $0.0035 \pm 0.0007$  &  $8.5 \times 10^{-4} \pm 1.0 \times 10^{-4}$  &  $1.1 \times 10^{-4} \pm 1.4 \times 10^{-4}$  &  $2.0 \times 10^{-5} \pm 8 \times 10^{-6}$  &  $ 6.1 \times 10^{-6} \pm 1.1 \times 10^{-6}$ \\
\hline
\end{tabular}
}
\caption{Probability of bit error (Mean $\pm$ stdev) as a function of the SNR (dB) for various conventional and nature-inspired multiuser detectors. 50 independent runs. Near-far effect: $E_k/E_1$=4 dB for $k \geq 1$.}
\label{tabla_m}
\end{table*}
%

Table \ref{tabla_m} shows the mean values and the corresponding standard deviations (SD), associated to Fig. \ref{fig_ber_snr1} plots. It is observed that  SD values decrease as SNR becomes higher, and  algorithms showing better convergence have a lower SD. Table \ref{tabla_m2} shows the best (minimum), mean, worst (maximum) and SD values for {\em SNR} $= 11$ and 14 dB cases.

\begin{table*}[htb]
\centering
\resizebox{15cm}{!}{
\begin{tabular}{|c|cccc|cccc|}
\hline
      &    \multicolumn{8}{c|}{ $SNR_{user \hspace{0.1cm} 1}$ (dB)}   \\
\hline
      &    \multicolumn{4}{c|}{ 11 }     &    \multicolumn{4}{c|}{ 14 }    \\
\hline
                   &  Best    & Mean  &  Worst &  SD &    Best     &    Mean  &  Worst     &    SD  \\
\hline
 Standard GA       &   $2.54 \times 10^{-3}$   &  $5.01 \times 10^{-3}$    &   $7.88 \times 10^{-3}$  &   $0.1 \times 10^{-2}$   &  $2.98 \times 10^{-4}$   &  $2.91\times 10^{-3}$   &   $4.98 \times 10^{-3}$  &  $0.9 \times 10^{-3}$ \\
\hline
Matched Filter     &   $7.18 \times 10^{-3}$   &  $1.10 \times 10^{-2}$   &  $1.59 \times 10^{-2}$   &  $0.2 \times 10^{-2}$    &  $5.92 \times 10^{-3}$   &   $9.65\times 10^{-3}$  &  $1.23 \times 10^{-2}$  &  $1.1 \times 10^{-3}$ \\
\hline
Decorrelator RLS   &   $1.01 \times 10^{-4}$   &  $2.12\times 10^{-3}$   &   $3.92 \times 10^{-3}$  &   $0.1 \times 10^{-2}$   &   $7.73 \times 10^{-5}$  &   $1.92\times 10^{-4}$    &   $3.25 \times 10^{-4}$  &  $7.0 \times 10^{-5}$ \\
 \hline
 Mahalanobis RBF   &   $1.11 \times 10^{-4}$   &  $6.31\times 10^{-4}$    &   $1.13 \times 10^{-3}$  &  $2.0\times 10^{-4}$   &     $1.46 \times 10^{-5}$ &   $1.95\times 10^{-5}$   &   $2.32 \times 10^{-5}$  &  $0.18 \times 10^{-5}$ \\
\hline
 MMSE-RLS          &    $5.63 \times 10^{-4}$  &  $8.51\times 10^{-4}$   &   $1.06 \times 10^{-3}$  &   $1.0\times 10^{-4}$  &     $1.52 \times 10^{-5}$ &    $6.52\times 10^{-5}$   &  $1.12 \times 10^{-4}$   &  $ 2.0 \times 10^{-5}$ \\
 \hline 
 Proposed FPA      &    $2.15 \times 10^{-4}$  &  $5.25\times 10^{-4}$  &   $7.98 \times 10^{-4}$  &   $1.3\times 10^{-4}$  &     $5.42 \times 10^{-6}$ &    $7.22\times 10^{-6}$  &  $9.33 \times 10^{-6}$   &  $0.9 \times 10^{-6}$ \\
  \hline 
 Tabu Search       &    $1.11 \times 10^{-4}$  &  $1.32\times 10^{-3}$   &  $2.63 \times 10^{-3}$   &   $6.0\times 10^{-4}$  &     $2.86 \times 10^{-6}$ &  $6.12\times 10^{-6}$   &  $9.69 \times 10^{-6}$   &  $1.4 \times 10^{-6}$ \\
 \hline
 Simulated Quenching &  $6.72 \times 10^{-4}$  &  $8.50\times 10^{-4}$   &   $1.08 \times 10^{-3}$  &   $1.0\times 10^{-4}$  &     $3.81 \times 10^{-6}$ &    $6.05\times 10^{-6}$   &  $8.74 \times 10^{-6}$   &  $1.1 \times 10^{-6}$ \\
 \hline
\end{tabular}
}
\caption{Best (minimum), mean, worst (maximum) and SD values of the Probability of Bit Error (BER) for different conventional and nature-inspired multiuser detectors. {\em SNR} $= 11$ and 14 dB cases. Results after 50 independent runs.}
\label{tabla_m2}
\end{table*}
%

Comparing the proposed FPA to a standard GA-based MUD, it can be seen how the FPA performance is close to the theoretical single user limit --notice that a lower limit for a multiuser detector algorithm is obtained simulating the probability of error when $U=1$ (absence of interfering users) \citep{Proakis98}--  while the GA-based MUD has difficulties to converge as SNR increases, even with a higher number of iterations and generations. This situation was already observed by \cite{Shayesteh03}.

Comparison with other nature-inspired algorithms such as simulated quenching (SQ) and tabu search (TS) show similar results whenever user 1 (UOI) transmits with high enough power (${\mbox{SNR}}_{\mbox{{\small user 1}}} \gtrsim 12$ dB) --see Fig. \ref{fig_ber_snr1}. In order to get a fair comparison, SQ and TS were implemented with an equivalent computational load measured in terms of fitness function evaluations. This is achieved by properly setting the number of iterations. Results are mean values after 50 independent runs. When ${\mbox{SNR}}_{\mbox{{\small user 1}}} < 12$ dB, both SQ and TS fail to achieve  acceptable estimates, while FPA still works correctly. The reason for this is that FPA combines more efficiently the explorative/exploitative sense of search, in part thanks to the adaptation of the probability of change, which controls local/global pollination, by monitoring the entropy of the population fitness, as explained in section \ref{sec_divmon}.

Performances of some traditional multiuser detectors, such as Matched Filter, Minimum Mean Square Error Estimator and  decorrelator detectors, are also shown in Fig. \ref{fig_ber_snr1}. Results indicate that simple algorithms (e.g. standard GA and MF) fail to converge to proper data estimates even when user 1 transmits with high power. On the other hand, nonlinear algorithms (RBF, MMSE and decorrelator) lead to an intermediate BER between SQ, TS and FPA, and that achieved with MF or standard GA.

Finally we compared the computation time of the proposed FPA with those of the following methods: Std-GA, TS and SQ. Every algorithm is iterated until the same bit error probability (BER) is achieved. Results are shown in Table \ref{table_time}. Table cells show percentage values normalized with respect to the time required by proposed FPA, whose time is shown as $100 \%$.

\begin{table}[htb]
\centering
\begin{tabular}{|c|c|c|c|}
\hline
         \multicolumn{4}{|c|}{ Processing time}   \\
\hline
   Proposed FPA   &    Std-GA    &  TS  &  SQ   \\
\hline
   100 $\%$       &  445 $\%$   & 150 $\%$   &  155 $\%$  \\
\hline
\end{tabular}
\caption{Comparison of processing time required to achieve the same BER. Percentage values normalized with respect to the time required by proposed FPA.}
\label{table_time}
\end{table}

We see that Std-GA requires about $4 \times$ time to achieve the same BER, while TS and SQ are in the order of $1.5 \times$.

\subsection{Statistical tests}

In order to evaluate the statistical significance of results involving nature-inspired algorithms, we performed a Kruskal-Wallis test with a Bonferroni correction. This test establishes if performance differences between Std-GA, SQ or TS, compared to the proposed FPA, are statistically significant. Table \ref{tabla_sigest} shows the results when a significance level $\alpha=0.05$ is used.

\begin{table}[htb]
\centering
\resizebox{8.5cm}{!}{
\begin{tabular}{|c|c|c|c|c|c|c|c|c|}
\hline
      &    \multicolumn{8}{c|}{ $SNR_{user \hspace{0.1cm} 1}$ (dB)}   \\
\hline
      &  7    & 8  &  9 &  10 & 11  & 12  &  13  & 14  \\
\hline
 Standard GA       &  $\dag$   & $\dag$   &  $\Box$  &  $\Box$  &  $\Box$  &  $\Box$  &  $\Box$  &  $\Box$ \\
\hline
Tabu Search  &  $\Box$   & $\Box$  &  $\Box$  &  $\Box$  &  $\Box$  &  $\dag$  &  $\Box$  &  $\dag$ \\
 \hline
 Simulated Quenching   &  $\Box$   & $\Box$  &  $\Box$  &  $\Box$  &  $\Box$  &  $\Box$  &  $\Box$  &  $\dag$ \\
\hline
\end{tabular}
}
\caption{Statistical significance of results using a Kruskal-Wallis test with Bonferroni correction ($\alpha = 0.05$). Each row compares proposed FPA against Std-GA, TS or SQ, respectively. $\Box$: FPA wins, $\dag$: difference is not statistically significant, $\diamond$: Std-GA/TS/SQ wins.}
\label{tabla_sigest}
\end{table}

In accordance to \citep*{Crepinsek14}, the algorithms were configured so as to have equal computational load in terms of fitness function evaluations. In particular, the number of generations is adjusted in order to have the same number of fitness function evaluations than FPA for each cell of Table \ref{tabla_sigest}.

Next, we compared performances of four metaheuristics using the Friedman test \citep{Friedman40}. The proposed algorithm is compared with Std-GA, TS and SQ. The problem test suite consists of 8 optimization problems, each one defined with a specific value of $\emph{SNR}_{\emph user 1}(dB) = \gamma$ dB, $\gamma \in [7,14]$. The Friedman test is conducted to detect significant behaviour differences between two or more detectors. Overage rankings are shown in Table \ref{tabla_X}. The best average ranking (lowest value) is in italics and corresponds to the FPA approach, which outperforms the other three schemes. A $p$-value of 0.0091 is obtained showing that there exist significant differences between the behaviour of the four algorithms.

\begin{table}[htb]
\centering
\begin{tabular}{|c|c|}
\hline
  Algorithm     &    Friedman Test Score   \\
\hline
FPA             &   {\em 1.625}   \\
SQ              &        1.875    \\
TS              &        3.00    \\
Std-GA          &        3.50    \\
\hline
\end{tabular}
\caption{Average ranking achieved with the Friedman test for four nature-inspired multiuser detectors: proposed FPA, standard GA, Tabu Search and Simulated Quenching. Each test problem comes given by a specific $\emph{SNR}_{\emph user 1}(dB)$ value, from set $\{7,8,9,10,11,12,13,14\}$.}
\label{tabla_X}
\end{table}

\subsection{Bit Error Rate {\it  vs} number of active users (capacity)}

\begin{figure}[htb] 
\centerline{\psfig{file=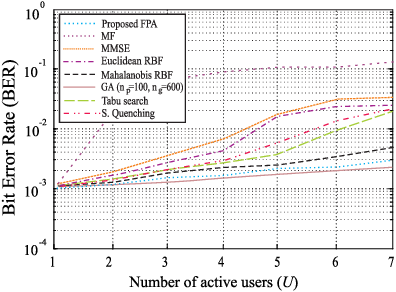,width=8cm}}
\caption{Bit Error Rate {\it vs} number of active users $U$. $E_i = E_j, 1 \leq  i,j  \leq U, i \neq  j$. GA is implemented with $n_p=100$ individuals and $n_g=600$ iterations. Average values after 30 independent runs.}   
\label{fig_ber_k}
\end{figure}

An important parameter used to assess the quality of a multiuser detector is its {\it capacity}. Capacity plots show the probability of erroneous symbol detection depending on the number of active system users. Numerical results are plotted in Fig. \ref{fig_ber_k}, where the error probabilities of several detectors (FPA, RBF, GA, SQ, TS, MMSEE and MF) are compared.

In this simulation  no near-far effect is present ($E_i = E_j, 1 \leq  i,j  \leq U, i \neq  j$). It can be seen how, on the basis of the BER metric, MF and MMSE methods are clearly outperformed by Mahalanobis-RBF and nature-inspired schemes. Under these conditions, multi-access interference greatly affects performance of MF, while the MMSE detector is not able to establish a good enough decision boundary. Notice that RBF implemented with the Mahalanobis distance yields better results than with the Euclidean one. 

Performances of nonlinear schemes  (the proposed FPA, Mahalanobis RBF and GA) are very close. However, it must be pointed out that both  RBF algorithms involve a much higher computational load (proportional to $2^U$) and require a supervised initial training period \citep{SanJoseNNSP03}.

On the other hand, we see that when the number of active users is moderate,  performances of GA, SQ, TS, Mahalanobis-RBF  and FPA  are, all of them, very similar. However, as the number of users increases, performance drops close to those of the MMSE and Euclidean-RBF schemes, since multi-access interference is more relevant.

It is obvious that performance degrades as $U$ increases, since population, in every evolutionary algorithm, is finite, and these methods are not able to properly explore the solutions space. A notable advantage of FPA and evolutionary approaches,  in contrast to conventional search methods, is that they do not need any memory elements, since information related to previous iterations does not need to be stored.

\subsection{Channel estimation performance}

\begin{figure}[htb]  
\centerline{\psfig{file=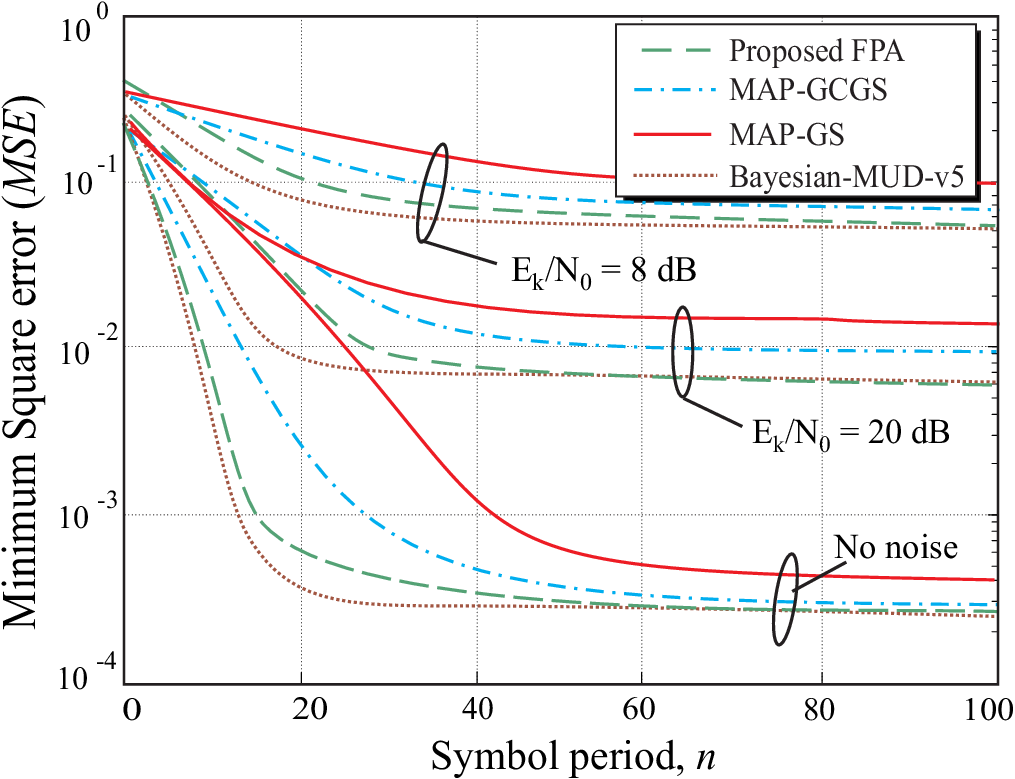,width=8cm}}
\caption{Average mean squared channel estimation error over a period of 100 symbols. The energy received for every active user is the same. Averaged values in 75 runs.}   
\label{f5y}
\end{figure}

The accuracy of fading coefficients' estimates is next analyzed. Performance is evaluated using the mean squared error (MSE) of the solution represented by the fittest flower when FPA finishes (${\bf b}^*$). Fig. \ref{f5y} shows comparative results with three different detectors: a Bayesian MUD proposed by \cite{SanJose07} and the so-called ``MAP-GCGS'' and ``MAP-GS'' algorithms developed by \cite{Huang02}, for several $E_k/N_0$ ratios measured along a 100-long symbols frame. The number of users is set to $U=8$, all of them transmitting with the same energy.

%

Numerical results show that, on average,  the proposed FPA requires about 30-40 symbol periods  to converge close to the final value, while MAP-GS and MAP-GCGS require about  40-60 samples, and the Bayesian-MUD-v5 detector takes a shorter period of about 25-30 samples.

It can be noticed that both Bayesian-MUD-v5 and FPA achieve a lower mean square error. However, the Bayesian scheme does not explicitly estimate the channel response coefficients, which the FPA method does. Besides, another advantage of nature-inspired methods like FPA is that they do not require any supervised initial period at all. Conversely, both MAP-GS and MAP-GCGS must be implemented with a 18 samples length burn-in period.

\begin{figure}[htb]  
\centerline{\psfig{file=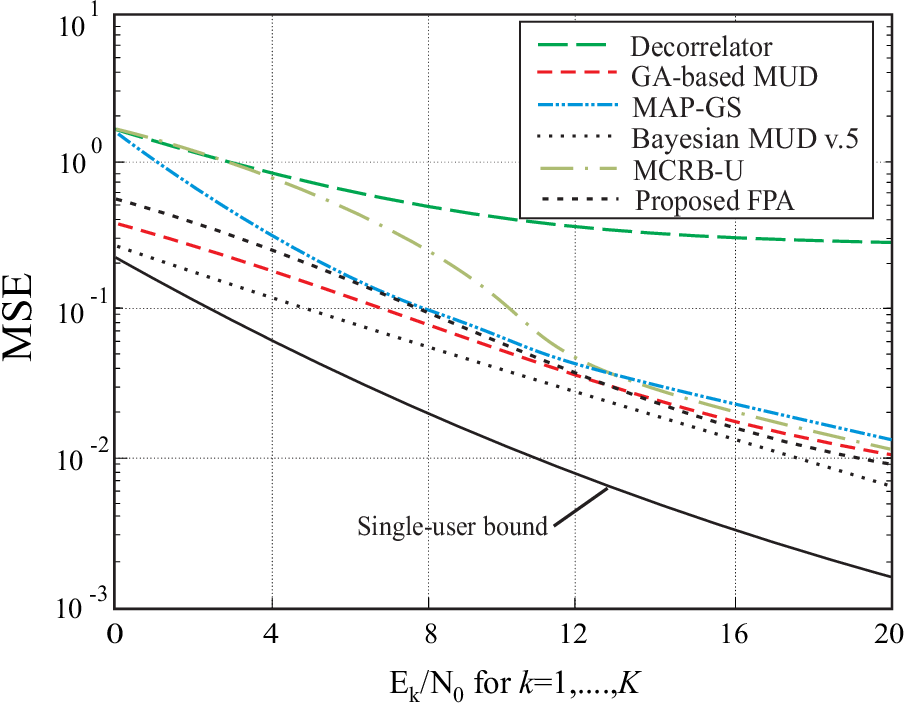,width=8cm}}
\caption{Average mean squared channel estimation error as a function of the users' bit energy. $U=8$. $E_i = E_j$, $i \neq j$, $1 \leq i,j \leq 8$. Every active user transmits with the same power.} 
\label{f7y}
\end{figure}

Next, Fig.\ref{f7y} compares the Mean Square Error (MSE) of the proposed method with five different detectors:  the conventional decorrelator estimator \citep{Lim98}, the Bayesian MUD v.5 \citep{SanJose07}, the MAP-GS algorithm \citep{Huang02}, the MCRB-U algorithm \citep{Farhang06} and the GA-based detector \citep{Yen01}. Error is represented as a function of the normalized symbol energy of the users ($E_i = E_j$, $i \neq j$, $1 \leq i,j \leq U$).  The single-user limit is plotted in black as a reference, as well.

Results show how nonlinear estimators, such as the GA-based MUD, the Bayesian MUD v.5 and the proposed FPA, present an importantly lower MSE than those of the decorrelator detector and the Bayesian approaches used for comparison (MAP-GS and MCRB-U), with a performance near the single-user limit, particularly for low and moderate SNRs ($< 11$ dB). Performances of FPA and Bayesian schemes are very similar when $E_k/N_0 \geq 12$ dB. However, for low SNR values, on the basis of the MSE metric, Bayesian MUD v.5, GA and FPA detectors outperform the other algorithms.

\subsection{Near-far performance}
\label{nearfar}

\begin{figure}[h]  
    \centerline{\psfig{file=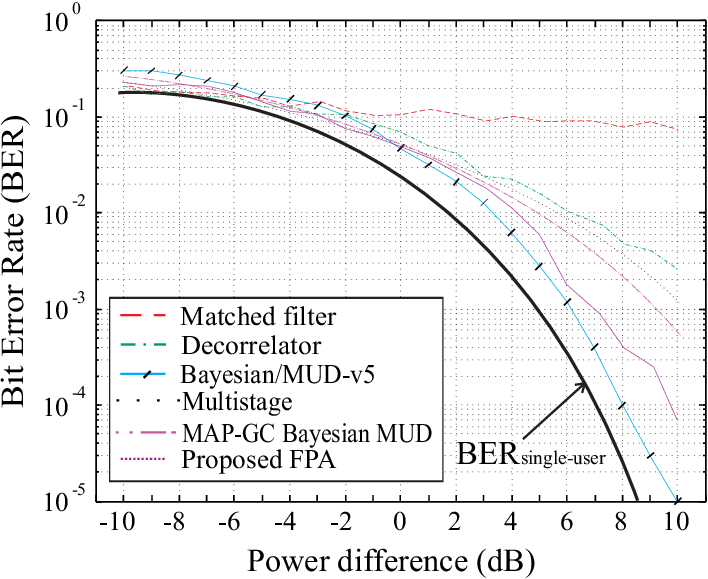,width=8cm}}
    \caption{Bit Error Rate of the user of interest (UOI) vs SNR for different multiuser detectors. Average results after 30 independent runs.}  
    \label{fig_nearfarsp8}
\end{figure}

The near-far effect considers the situation in which  part of the  interfering users are closer to the base-station than the user of interest (UOI) is. This involves the reception of stronger undesired signals compared to the one coming from the user of interest, making, this way, more difficult the problem of satisfactorily extracting useful data \citep{Proakis98,Verdu98}. Bit Error Rate of the UOI as a function of the difference of received power between the UOI and the other users is evaluated. Three interfering users, i.e. $U=4$, processing gain  $N=32$ and noise variance $\sigma_n^2 = 0.5$, are considered. Besides,  powers from the three interfering users are the same at the receiver antenna. In Fig. \ref{fig_nearfarsp8} the BER obtained for the UOI versus $E_1/E_k$ ($2 \leq k \leq U$) is shown  for several detectors. Error probabilities of two conventional detectors --the decorrelator \citep{Lim98} and the MF scheme \citep{Lupas89}-- are shown for comparison, as well as the performance of a multistage detector \citep{Buehrer96} and the MAP-GC Bayesian scheme \citep{Huang02}.

For this simulation, intersymbol interference is assumed by considering that the transfer function of the UOIs channel response is $H_{\mbox{UOI}}(z) = a_0 + 0.471z^{-1} + 0.063z^{-2}$. The first coefficient ($a_0$) varies in such a way that signal-to-in\-ter\-fer\-ence ratio is $-10 \leq \mbox{SIR} \leq 10$ dB. We can see how BER decreases as the power difference increases, what\-ev\-er detector scheme is used, which makes sense, since the received signal from the UOI becomes stronger at the receiver as both $a_0$ and the power difference with respect to interferers, increase.

Results show that both Bayesian-MUD-v5 \citep{SanJose07} and the proposed FPA are near-far resistant, achieving a bit error rate close to the single user limit, whilst   MF shows a notable degradation --it is well-known that interferences limit its performance. On the other hand, on the basis of the BER metric, the decorrelator detector is somewhat outperformed by the MAP-GC algorithm and mul\-ti\-stage detectors.

 If computational load is analyzed by fixing $U=4$, $\sigma_n^2 = 0.5$ 
 and a power difference of 2 dB, it can be seen that the decorrelator detector and the proposed FPA have a nice trade-off between complexity and performance (decorrelator takes about $1.3 \times$ the time of FPA). FPA shows good efficiency,  close to the Bayesian scheme, with only a small BER degradation, while demanding less computational load (the Bayesian scheme is $1.8 \times$ less efficient than FPA). On contrast, the MAP-GC algorithm is the most complex and time-consuming algorithm ($2.5 \times$ of FPA).

\begin{figure}[htb]  
    \centerline{\psfig{file=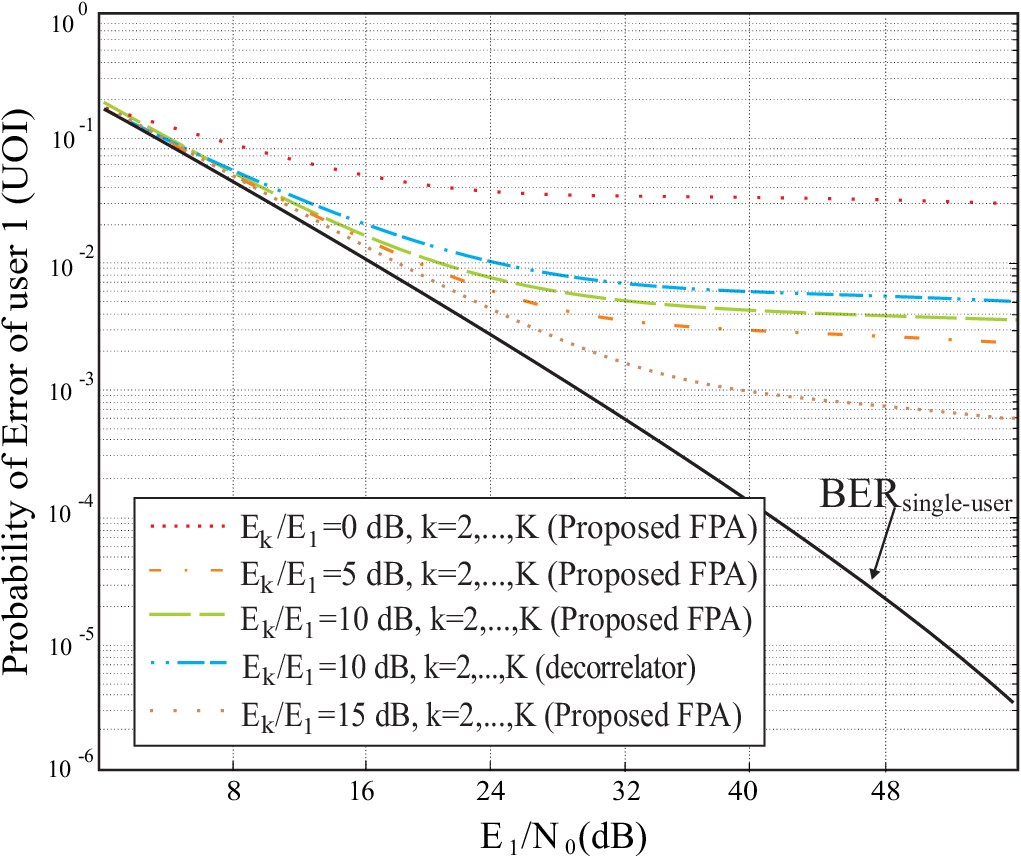,width=8cm}}
    \caption{Bit Error Rate performance for $K=10$ active users with $E_k/E_1=0$, 5, 10 and 15 dB for $k=2,\dots,U$. User 1: user of interest (UOI).}  
    \label{fig_nearfar2sp8}
\end{figure}

Finally, Fig. \ref{fig_nearfar2sp8} shows the near-far performance of the proposed FPA detector depending on the BER of the UOI. The averaged received symbol energies of the remaining users are adjusted 0, 5, 10 and 15 dB higher than the UOI's energy. Numerical results show that when $5 \leq E_k/E_1 \leq 10$ dB, the proposed FPA has good near-far capabilities up to $E_1/N_0 \approx 26$ dB. Higher values of $E_1/E_0$ lead to a slight BER performance degradation.

The BER plot corresponding to a conventional decorrelator detector is shown for $E_k/E_1=10$ dB to allow comparison, as well. It can be seen  how, on the basis of the probability of bit error of the UOI metric, performance drops a little with this method with respect to the FPA scheme.
\\
\\
Consequently, we can say that the proposed FPA  multiuser detector is near-far resistant and robust when both inter-symbol and multi-access interferences exist, constituting an interesting alternative to already existing methods.

\section{Conclusions}

In this work, a multiuser detector based on FPA has been developed to jointly solve the problems of symbol detection and channel estimation in a DS/CDMA multiuser communication environment, which is one of the most important and implemented channel sharing techniques in advanced communication systems, such as some parts of LTE and 5G standards.

Among the improvements introduced over the standard FPA, the following can be highlighted:  scale factors associated with  Levy flight and change probability are adjusted according to the entropy of the aptitude of the flowers so that an optimal compromise between exploitation and exploration is kept. Besides, two different parts are considered within the encoding of each flower, and the pollination processes are applied differently to each part.

The algorithm obtained in this way is implementable in real CDMA systems (in LTE, GPS, IS95, 5G, etc.), allowing to work in more critical conditions (lower received power, greater interferences, large fading, higher number of users) and at higher transmission rates, while keeping conceptual and computational simplicity.

Numerical simulations have been carried out, as far as we know, in fair terms, including comparisons with several powerful and well-known conventional schemes as well as with other metaheuristics. The proposed method is proved to be an efficient tool to solve the complex problem of equalization in digital communications. FPA offers competitive performance, especially when conditions become more difficult (a large number of interfering users, low SNR for the user of interest or the existence of near-far effects) and less complexity than that of conventional methods achieving similar  performance. When inter-symbol and multiple-access interferences are not high, other nature-inspired schemes, such as TS, GA and SQ, show similar performance to those of the proposed detector and the optimal receiver. But when  interferences become stronger, or the power of the user of interest is lower, the proposed FPA performs better, in part due to its powerful search capabilities that adjust the diversity of the population by monitoring the entropy of the population fitness. This way, the probability of change, which selects between  local and global pollination, is in-service adjusted using this information. Finally, several tests have been run to analyze the statistical significance of results.

Among the limitations of our work, we highlight the difficulty and dependence of the optimal initial adjustment of  parameters. Besides, the paper is  focused on a binary source alphabet and considers a perfectly synchronous system.

As future lines of work we consider the extension to non-binary modulations, asynchronous systems and other channel sharing schemes. Likewise, our next works will be aimed at the detailed study of the influence of each parameter of the algorithm, a more detailed study of the computational load and a deeper statistical analysis of the results.


\appendix

\section{Particularization to a chip-rate algorithm}
\label{Apendice_A}

In this case, taking into account that the energy of the chips is $\mathcal{E}_\gamma = \int_0^{T_c} |\gamma(t)|^2 dt$, a sequence of $N$ samples is obtained for each symbol, whose components are calculated as
\begin{multline}
     r_{n,j} = \int_{nT+jT_c}^{nT+(j+1)T_c} r(t) \gamma(t-nT-jT_c) dt   \\
            =    \sum_{u=1}^U a_u(n) d_u(n) s_u(j) \mathcal{E}_\gamma +  \int_0^{T_c} g(t+nT-jT_c) \gamma(t) dt, \\
            \hspace{2cm} j=0,1,\dots,N-1
\end{multline}
This set of $N$ samples is normalized by the chip energy $\mathcal{E}_\gamma$ and grouped into a vector ${\bf r}_n^{chip}$. Thus, we can write
\begin{equation}
     {\bf r}_n^{chip}  = \sum_{i=1}^U {\bf s}_i a_i(n) d_i(n) + {\bf g}(n) = {\bf SA}{\bf d}(n) + {\bf g}(n)
\label{rnchip}
\end{equation}
where ${\bf S}=[{\bf s}_1,{\bf s}_2,\dots,{\bf s}_U]$ is the $N\times U$ matrix whose columns contain the users' signatures ${\bf s}_i$, and ${\bf g}(n) = [g_{n,0}, g_{n,1},\dots, g_{n,L-1}]^T$ stands for the normalized noise vector with components
\begin{equation}
     g_{n,j} = \frac{1}{\mathcal{E}_\gamma} \int_0^{T_c} g(t+nT+j T_c) \gamma(t) dt, \qquad j=0,1,\dots,N-1
\end{equation}
Since {\bf g} represents a zero-mean, white and Gaussian noise process, its covariance matrix is $E\{ {\bf g}(n) {\bf g}(n)^H \}  = \sigma^2 {\bf I}_N$, with ${\bf I}_N$ being the $N \times N$ identity matrix.

\section{Abbreviations}
\label{Apendice_B}

The main acronyms used in the paper are:
\\
\\
ACO: Ant colony algorithm \\
BPSK: Binary Phase Shift Keying \\
CSO: Cat swarm optimization \\
DS/CDMA: Direct Sequence / Code-Division Multiple-Access \\
FPA: Flower pollination algorithm \\
GA: Genetic algorithm \\
GPS: Global positioning system \\
ISI: Intersymbol interference \\
MAI: Multiaccess interference \\
MF: Matched filter \\
ML: Maximum Likelihood \\
MMSEE: Minimum Mean Square Error Estimator \\
MUD: Multiuser detector \\
PSO: Particle swarm optimization \\
RBF: Radial basis function \\
SA: Simulated annealing \\
SNR: Signal-to-noise ratio \\
SQ: Simulated quenching \\
TS: Tabu search  \\
UOI: User of interest \\


%
%


\subsection*{{\bf Compliance with ethical standards}}

{\bf Conflict of Interest} The authors declare that they have no conflict of interest. \\ \\
{\bf Ethical approval} This article does not contain any studies with human participants or animals performed by any of the authors.

\bibliographystyle{model5-names}  
\bibliography{refs}   

%
%

\end{document}